\documentclass{aa} 

\usepackage{graphicx}
\usepackage[english]{babel}
\usepackage{natbib}
\usepackage[pdftex]{hyperref}
\usepackage{amsmath} 
\usepackage{txfonts}

\bibpunct{(}{)}{;}{a}{}{,}

\begin{document}

\title{Population synthesis of classical low-mass X-ray binaries\\ in the Galactic Bulge} 

\author{
L.~M.~van~Haaften\inst{\ref{radboud},\ref{ttu}} \and
G.~Nelemans\inst{\ref{radboud},\ref{leuven}} \and
R.~Voss\inst{\ref{radboud}} \and
M.~V.~van~der~Sluys\inst{\ref{radboud}} \and
S.~Toonen\inst{\ref{leiden},\ref{radboud}}
}

\institute{
Department of Astrophysics/ IMAPP, Radboud University Nijmegen, P.O. Box 9010, 6500 GL Nijmegen, The Netherlands, \email{L.vanHaaften@ttu.edu} \label{radboud} \and
Department of Physics, Texas Tech University, Box 41051, Lubbock TX 79409-1051, USA \label{ttu} \and
Institute for Astronomy, KU Leuven, Celestijnenlaan 200D, 3001 Leuven, Belgium \label{leuven} \and
Leiden Observatory, Leiden University, P.O. Box 9513, 2300 RA Leiden, The Netherlands \label{leiden}
}

\abstract {} 
{We model the present-day population of classical low-mass X-ray binaries (LMXBs) with neutron star accretors, which have hydrogen-rich donor stars. Their population is compared with that of hydrogen-deficient LMXBs, known as ultracompact X-ray binaries (UCXBs). We model the observable LMXB population and compare it to observations. We model the Galactic Bulge because it contains a well-observed population and it is the target of the Galactic Bulge Survey.} 
{We combine the binary population synthesis code \textsf{SeBa} with detailed LMXB evolutionary tracks to model the size and properties of the present-day LMXB population in the Galactic Bulge. Whether sources are persistent or transient, and what their instantaneous X-ray luminosities are, is predicted using the thermal-viscous disk instability model.} 
{We find a population of $\sim \! 2.1 \times 10^{3}$ LMXBs with neutron star accretors. Of these about $15 - 40$ are expected to be persistent (depending on model assumptions), with luminosities higher than $10^{35}\ \mbox{erg s}^{-1}$. About $7 - 20$ transient sources are expected to be in outburst at any given time. Within a factor of two these numbers are consistent with the observed population of bright LMXBs in the Bulge. This gives credence to our prediction of the existence of a population of $\sim \! 1.6 \times 10^{3}$ LMXBs with low donor masses that have gone through the period minimum, and have present-day mass transfer rates below $10^{-11}\ M_{\odot}\ \mathrm{yr}^{-1}$.} 
{Even though the observed population of hydrogen-rich LMXBs in the Bulge is larger than the observed population of (hydrogen-deficient) UCXBs, the latter have a higher formation rate. While UCXBs may dominate the total LMXB population at the present time, the majority would be very faint or may have become detached and produced millisecond radio pulsars. In that case UCXBs would contribute significantly more to the formation of millisecond radio pulsars than hydrogen-rich LMXBs.} 

\authorrunning{L.~M.~van~Haaften et al.}
\titlerunning{Population synthesis of LMXBs in the Galactic Bulge}

\maketitle

\section{Introduction}
\label{sect:lmxb:intro}

Low-mass X-ray binaries (LMXBs) are binaries in which a star with a mass below $\sim \! 1.5\ M_{\odot}$ transfers mass via Roche-lobe overflow to a neutron star or black hole companion \citep[e.g.,][]{vandenheuvel1975,joss1979}.
The bifurcation period separates zero-age main-sequence LMXBs with expanding orbits and long orbital periods, in which mass transfer is driven by donor evolution, from LMXBs with shrinking orbits, in which mass transfer is driven by angular momentum loss via magnetic braking and gravitational wave radiation \citep{tutukov1985,pylyser1988,pylyser1989}. 
In the first group of `diverging', long-period systems, the donor star will eventually become a detached white dwarf after its subgiant or giant progenitor has lost almost all of its envelope \citep{webbink1983}. In the `converging' systems, however, the donor loses most of its mass before it has developed a helium core. Convection turns the donor into a homogeneous mixture of hydrogen and helium, where the ratio between both elements depends on the evolutionary stage at the time hydrogen burning was extinguished \citep[e.g.,][]{tutukov1985}.

Similar to cataclysmic variables, the orbital periods of the bulk of the converging LMXBs shrink until a period minimum of $\sim \! 70 - 80$ min \citep{paczynski1981cv,paczynski1981sien,rappaport1982}, where the donor reaches a maximum average density, and later it becomes degenerate. In the case of sufficiently evolved cores, the period may continue to shrink to $\sim \! 40$ min \citep{nelson2003} or even $\sim \! 5 - 10$ min \citep{tutukov1985,podsiadlowski2002,sluys2005a}. Systems subsequently slowly expand towards longer orbital periods.
Stable mass transfer continues to decrease the donor mass, and irradiation of the donor by the accretor and the accretion disk becomes important in driving mass transfer and generating a wind from the donor \citep{ruderman1989late}.

In both the converging and diverging systems, neutron star accretors can be recycled to spin periods of a few milliseconds. If accretion stops, systems can become binary millisecond radio pulsars, as has been observed in the $4.75$ hr binary \object{FIRST J102347.67+003841.2} \citep{archibald2009}. The observed evaporation process suggests the possibility that `black widow' systems may turn into isolated millisecond radio pulsars \citep{ruderman1989,fruchter1988}.

In this study we estimate the number of classical (i.e., hydrogen-rich) LMXBs in the Galactic Bulge, and compare the results to observations of bright LMXBs, as well as with earlier population synthesis studies. Furthermore, we compare the results with the population of ultracompact X-ray binaries (UCXBs) in the Bulge modeled by \citet{vanhaaften2013bulgeucxb}, and to results by the Galactic Bulge Survey \citep{jonker2011,jonker2014}.

\section{Method}
\label{sect:lmxb:method}

We use the binary population synthesis code \textsf{SeBa} \citep{portegieszwart1996,nelemans2001a,toonen2012} to simulate the primordial Galactic Bulge binary population, and select all binaries containing a neutron star and a main-sequence star, immediately after the supernova event.
We use a combined common-envelope parameter $\alpha_\mathrm{CE} \lambda = 2$ for massive stars \citep[e.g.,][]{portegieszwart1998}. In Sect.~\ref{sect:lmxb:discussion} we also consider the effect of using a lower value.
The initial mass function of the primary components is derived by combining the initial mass function for stellar systems \citep{kroupa2001} with the binary fraction as a function of mass. The mass ratios of the secondary and primary components follow a uniform distribution between $0$ and $1$, and the eccentricity $e$ is drawn from a distribution proportional to $e$ between $0$ and $1$. The semi-major axis $a$ distribution is inversely proportional to $a$ \citep{popova1982}, with a lower limit determined by the stellar radii, and an upper limit of $10^{6}\ R_{\odot}$ \citep{duquennoy1991}. The kick velocities imparted on neutron stars during their formation are drawn from the distribution by \citet{paczynski1990} with a dispersion of $270\ \mbox{km s}^{-1}$.
For more details on the initial binary parameters, we refer to \citet{vanhaaften2013bulgeucxb}.

\begin{figure}
\resizebox{\hsize}{!}{\includegraphics{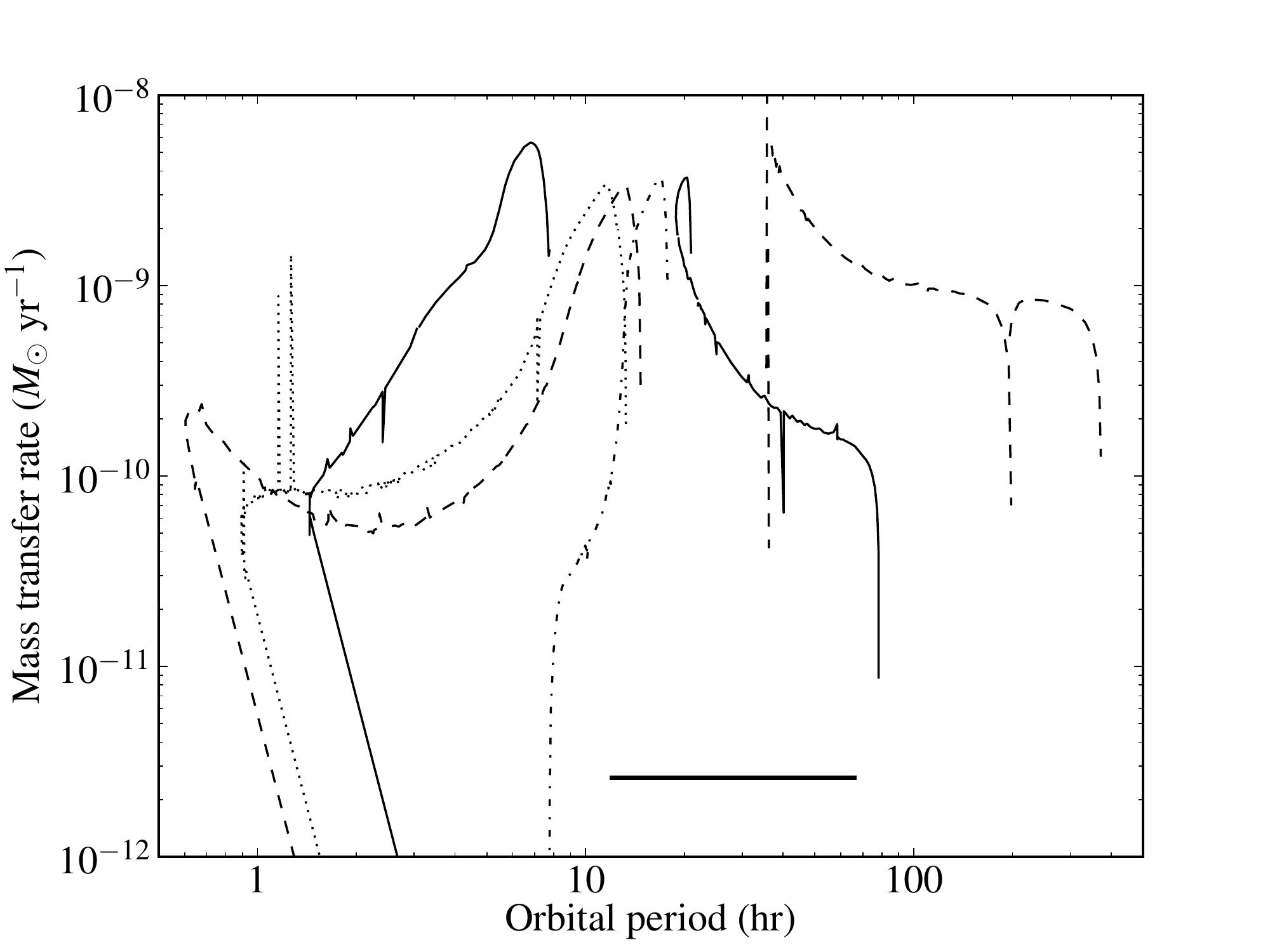}} 
\caption{Sample of tracks showing the mass transfer rate against orbital period for LMXBs with initial donor masses of $1\ M_{\odot}$. The complete grid covers a range of initial orbital periods between $0.5$ and $2.75$ days ($12 - 66$ hr, indicated by the horizontal bar) \citep{sluys2005a}.}
\label{fig:lmxb:permdot_m1}
\end{figure}

For the subsequent evolution of the selected systems we use the low-mass X-ray binary tracks described in \citet{sluys2005a}, a selection of which is shown in Fig.~\ref{fig:lmxb:permdot_m1}. These tracks describe the evolution of binaries starting with a neutron star accretor and a zero-age main-sequence donor. The tracks cover a grid of initial donor masses and orbital periods. The donor masses range from $0.7$ to $1.5\ M_{\odot}$ with steps of $0.1\ M_{\odot}$. The initial orbital periods range from $0.5$ to $2.75$ days, with steps of $0.25$ days, but with $0.05$ day steps near the bifurcation period.
The spikes in Fig.~\ref{fig:lmxb:permdot_m1} have a numerical origin; they are caused by the finite number of grid points in the stellar-structure models. When a convective region shrinks or expands, a grid point near the boundary will start or stop being convective. Hence, the changes in convective regions are not smooth, but undergo discrete steps. Such changes cause steps in the radius of the star, which influences its Roche-lobe-filling factor, to which the mass-transfer rate is rather sensitive, causing the spikes.
Each system produced by \textsf{SeBa} is matched to the track that most closely approaches its donor mass and orbital period. We correct the orbital period of the \textsf{SeBa} system because its donor is slightly evolved at the time of the supernova event, whereas the tracks start with zero-age main-sequence donors. 

In Fig.~\ref{fig:lmxb:permdot_m1}, the orbital periods have shrunk before the systems start mass transfer. The two tracks on the right-hand side show LMXBs that diverge, and ultimately become detached. The three tracks on the left-hand side represent converging systems, which go through a period minimum. Among the converging systems, the lowest period minimums are reached by systems with the longest initial period, as long it lies below the bifurcation period.

Initially the orbital evolution of the binary is predominantly driven by magnetic braking of the main-sequence star. We calculate the angular momentum loss rate using the formula by \citet{rappaport1983}.
A metallicity $Z = 0.02$ is used in \textsf{SeBa}. The LMXB tracks use $Z = 0.01$, but their evolution is very similar for $Z = 0.02$ \citep{sluys2005a}.

Following \citet{vanhaaften2013bulgeucxb}, we approximate the star formation history of the Galactic Bulge by a Gaussian distribution with a mean $\mu = -10$ Gyr (the present time is defined as $0$) and a standard deviation $\sigma$ of either $0.5$ or $2.5$ Gyr, in order to model both burst-like and extended star formation. For the total star forming mass we take $1 \times 10^{10}\ M_{\odot}$ \citep{clarkson2009,wyse2009}.

The X-ray luminosity $L_\mathrm{X}$ of a system is calculated from its mass transfer rate using
\begin{equation}
    \label{eq:lmxb:lum}
    L_\mathrm{X} = \eta_\mathrm{bol} \frac{GM_\mathrm{a}\dot{M}_\mathrm{a}}{R_\mathrm{a}},
\end{equation}
where $\eta_\mathrm{bol} \approx 0.55$ is the bolometric correction \citep[following][]{belczynski2008}, $M_\mathrm{a}$ is the accretor mass, $R_\mathrm{a}$ is the accretor radius, and $G$ the gravitational constant.
We use the thermal-viscous disk instability model \citep[see, e.g., the review by][]{lasota2001} to decide which of our modeled sources are transient and which are persistent at the present time. Sources do not have a stable disk if their mass transfer rates are below the critical value $\dot{M}_\mathrm{crit}$ for irradiated accretion disks derived by \citet{dubus1999}, in the form given by \citet{zand2007}
\begin{equation}
    \label{eq:lmxb:inst}
    \dot{M}_\mathrm{crit} \approx 5.3 \times 10^{-11}\ f \left( \frac{M_\mathrm{a}}{M_{\odot}} \right)^{0.3} \ \left( \frac{P_\mathrm{orb}}{\mathrm{hr}} \right)^{1.4}\ \ M_{\odot}\ \mathrm{yr}^{-1},
\end{equation}
where $P_\mathrm{orb}$ is the orbital period and $f$ is a scale factor depending on the disk composition; $f \approx 1$ for solar-composition disks and $f \approx 6$ for helium disks. Systems with unstable accretion disks are expected to be transient sources, which show outbursts separated by (much longer) quiescent phases. The luminosity during outbursts is given by Eq.~(\ref{eq:lmxb:lum}) with accretion rate $\dot{M}_\mathrm{a} = \dot{M}_\mathrm{crit}$. The luminosity is assumed to be zero during quiescence. Therefore, the duty cycle is equal to the ratio of the time-averaged mass transfer rate and the mass transfer rate during outburst. The time-averaged mass transfer rate is the rate at which the donor loses mass according to the evolutionary tracks. This rate is virtually constant on timescales much shorter than the evolutionary timescales of the donor and the binary system.
If sources have time-averaged mass transfer rates higher than the critical value, we assume they are persistent X-ray sources with X-ray luminosities given by Eq.~(\ref{eq:lmxb:lum}) where $\dot{M}_\mathrm{a}$ is taken to be the time-averaged mass transfer rate.

We do not model LMXBs with black hole accretors because it is not at all clear how they are formed \citep{portegieszwart1997,kalogera1999,podsiadlowski2003,justham2006,yungelson2006}, and in particular how the binary system remains bound during the formation of the black hole, given the low mass of the companion. Additionally, all observed (certain) black hole LMXBs are transient and therefore we cannot compare model results to persistent systems.

\section{Results}
\label{sect:lmxb:results}

\subsection{The total population of LMXBs}

The total number of LMXBs with neutron star accretors formed over the history of the Galactic Bulge in our study is $7.6 \times 10^{3}$. Of these, $(2.0 - 2.1) \times 10^{3}$ (for star formation history width $\sigma = 2.5$ and $0.5$ Gyr, respectively) are still X-ray binaries at the present time -- all other systems have stopped transferring mass and have turned into long-period detached white dwarf--neutron star binaries.

\begin{figure}
\resizebox{\hsize}{!}{\includegraphics{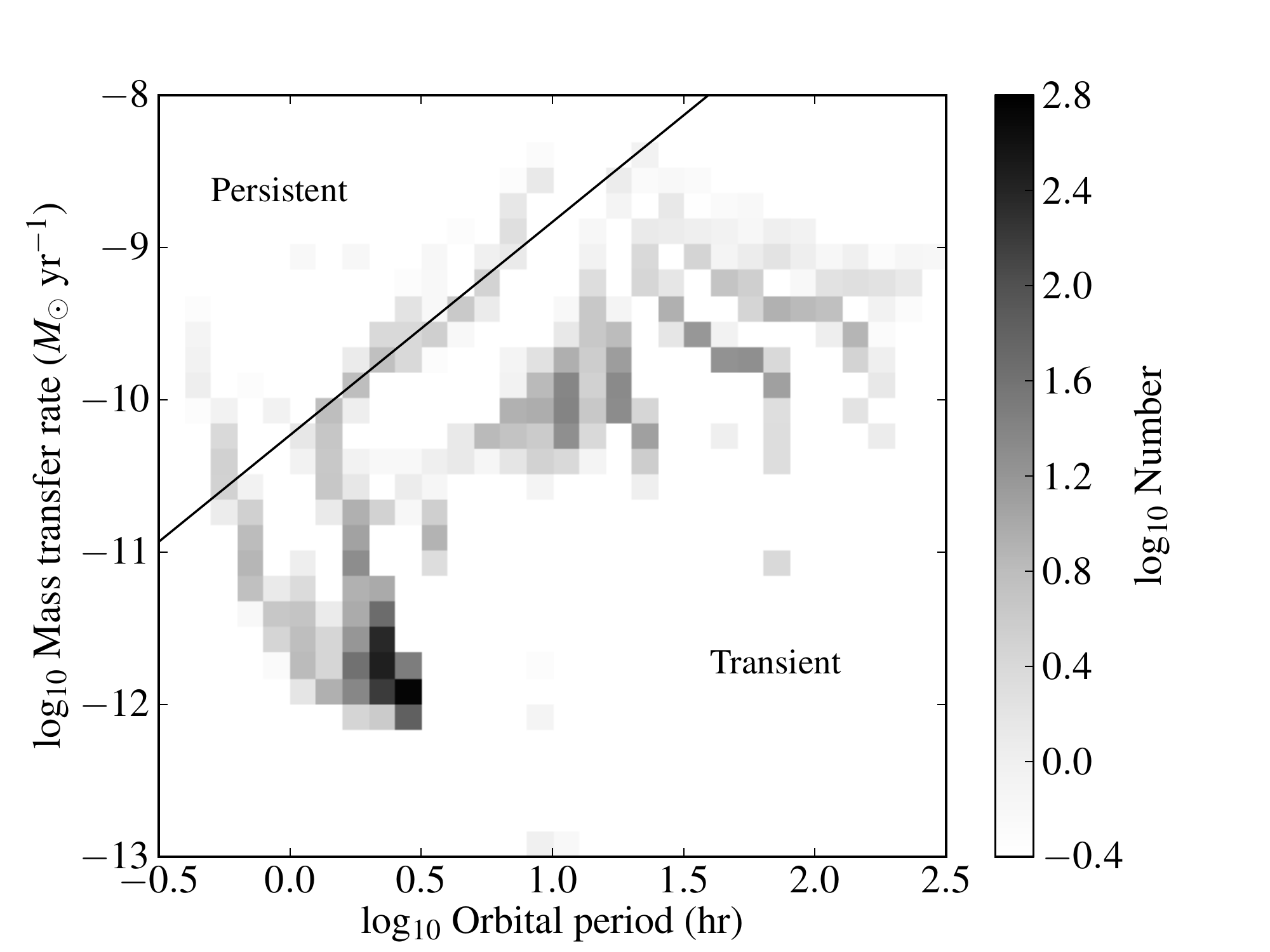}} 
\caption{Model distribution of present-day mass transfer rate versus orbital period for hydrogen-rich LMXBs in the Galactic Bulge. White squares correspond to $0.4$ or fewer LMXBs. The highest density in the figure is $10^{2.7} \approx 500$ per pixel. The black line shows the critical mass transfer rate for thermal-viscous disk instability, given by Eq.~(\ref{eq:lmxb:inst}) with $M_\mathrm{a} = 1.4\ M_{\odot}$ and $f = 1$ (solar composition). The star formation history width $\sigma = 0.5$ Gyr.}
\label{fig:lmxb:density05}
\end{figure}
\begin{figure}
\resizebox{\hsize}{!}{\includegraphics{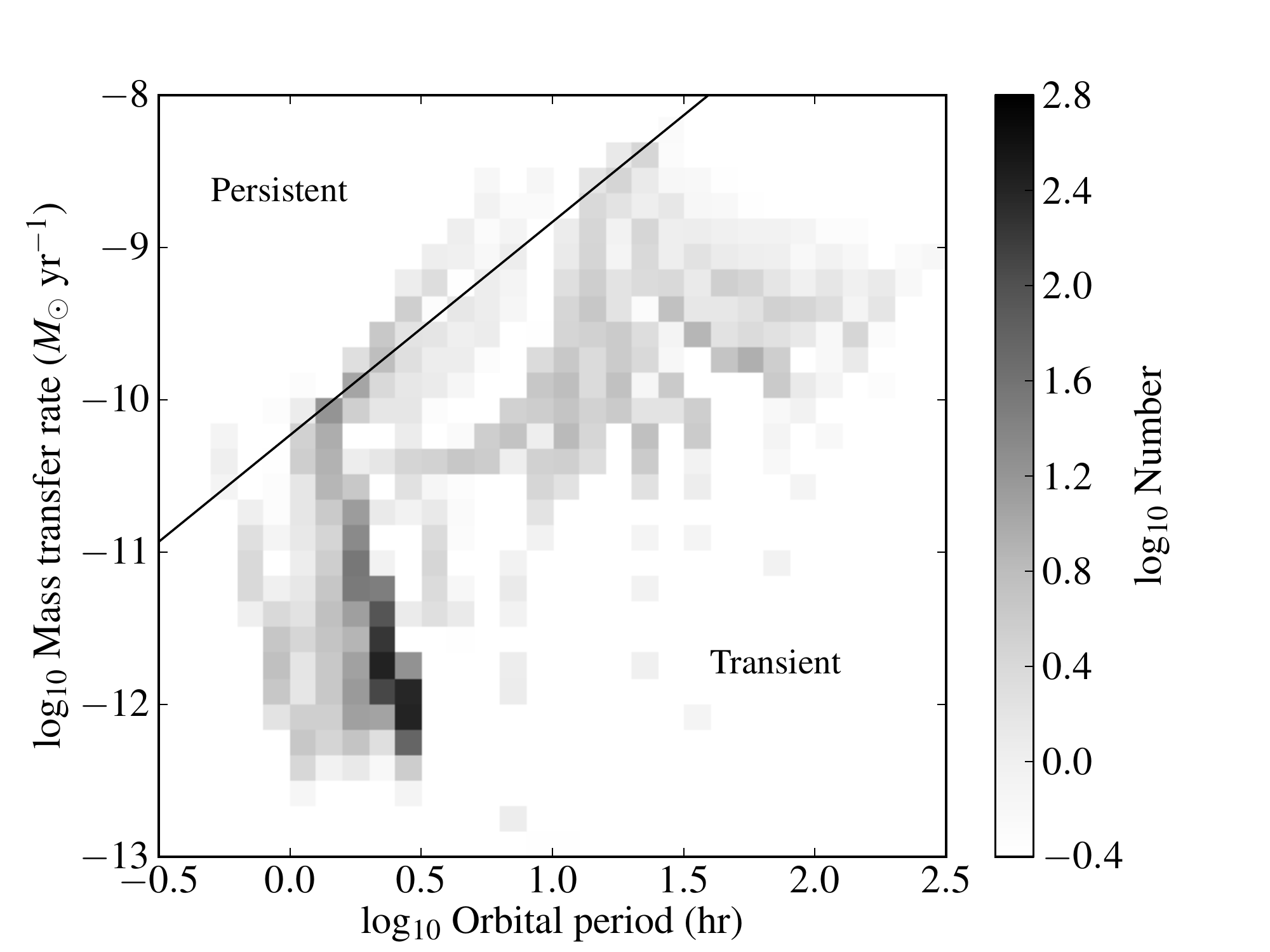}} 
\caption{Same as Fig.~\ref{fig:lmxb:density05}, except $\sigma = 2.5$ Gyr. The highest density in the figure is $10^{2.4} \approx 250$ per pixel.}
\label{fig:lmxb:density25}
\end{figure}

Figures~\ref{fig:lmxb:density05} and \ref{fig:lmxb:density25} show the present-day population of $\sim \! 2 \times 10^{3}$ LMXBs in terms of their orbital periods and time-averaged mass transfer rates, for the two star formation history widths (Sect.~\ref{sect:lmxb:method}). There are $(1.5 - 1.6) \times 10^{3}$ systems (for $\sigma = 2.5$ and $0.5$ Gyr, respectively), three quarters of all LMXBs, with mass transfer rates below $10^{-11}\ M_{\odot}\ \mathrm{yr}^{-1}$ and orbital periods shorter than $3$ hr. These LMXBs have passed through the orbital period minimum and have increasing periods at the present time. The donor stars in these systems are (semi-)degenerate and have masses below $\sim \! 0.05\ M_{\odot}$. The donor stars in systems that reach periods shorter than $\sim \! 60$ min have no hydrogen in their cores and less than $10\%$ hydrogen on their surface at the period minimum \citep{sluys2005a}.

Almost all LMXBs with orbital periods longer than $\sim \! 3$ hr have mass transfer rates above $\sim \! 10^{-11}\ M_{\odot}\ \mathrm{yr}^{-1}$, and about three quarters of these are diverging systems.
The solid lines in Figs.~\ref{fig:lmxb:density05} and \ref{fig:lmxb:density25} separate the persistent and transient sources according to the Disk Instability Model (Sect.~\ref{sect:lmxb:method}). In Sect.~\ref{sect:lmxb:observable} we will discuss this in more detail.


\begin{figure}
\resizebox{\hsize}{!}{\includegraphics{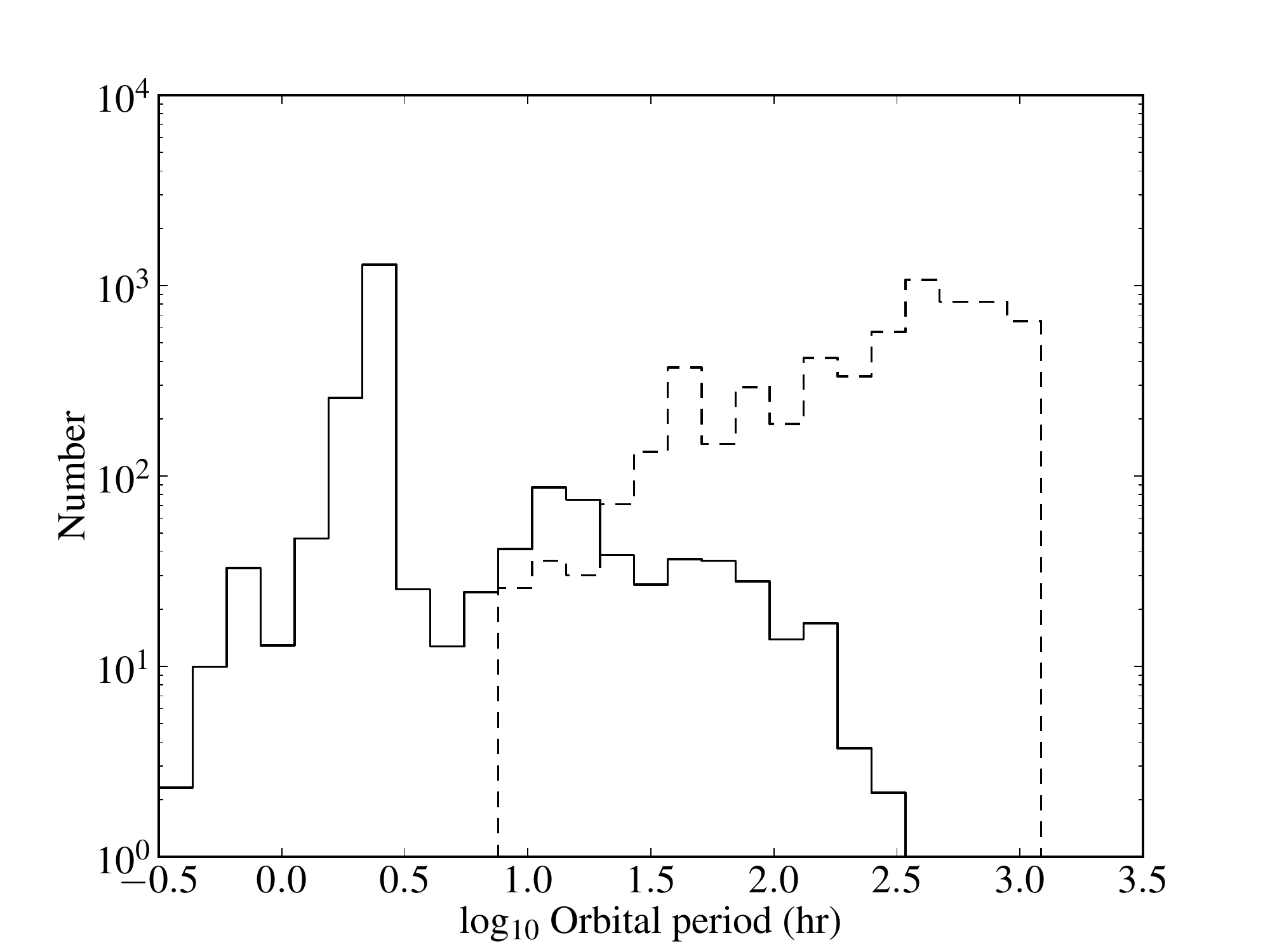}} 
\caption{Orbital period distributions of the present-day populations of hydrogen-rich LMXBs with neutron stars accretors (solid line), and LMXBs that have evolved into detached white dwarf--neutron star systems (dashed line) in the Galactic Bulge. The star formation history width $\sigma = 0.5$ Gyr.}
\label{fig:lmxb:perhist_log05}
\end{figure}
\begin{figure}
\resizebox{\hsize}{!}{\includegraphics{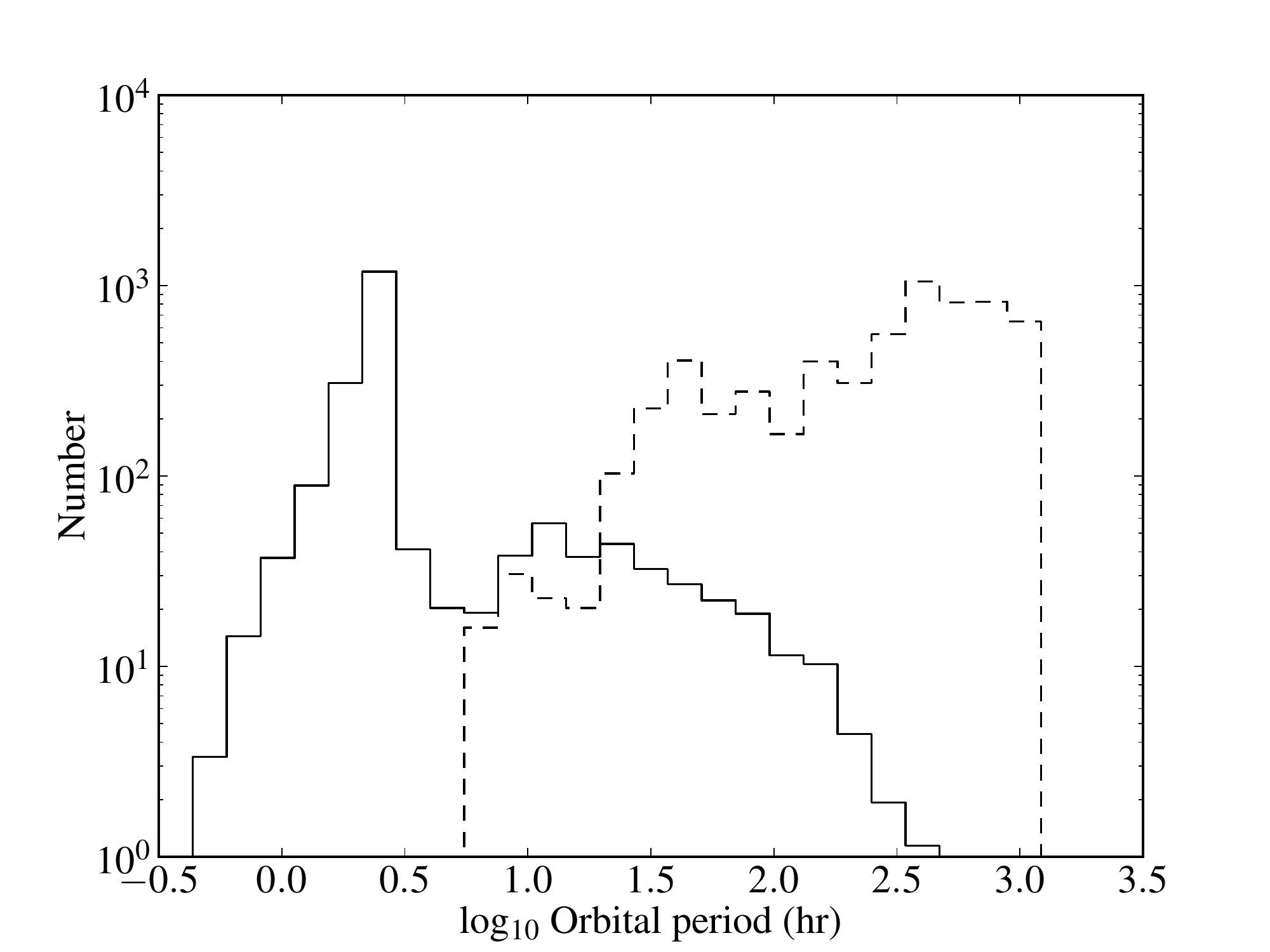}} 
\caption{Same as Fig.~\ref{fig:lmxb:perhist_log05}, except $\sigma = 2.5$ Gyr.}
\label{fig:lmxb:perhist_log25}
\end{figure}

The orbital period distributions of all present-day neutron star systems, both semi-detached and detached, are shown in Figs.~\ref{fig:lmxb:perhist_log05} and \ref{fig:lmxb:perhist_log25}, again for two star formation histories. The semi-detached systems (solid lines) are the same populations as shown in Figs.~\ref{fig:lmxb:density05} and \ref{fig:lmxb:density25}, combining all mass transfer rates. Of all LMXBs that have formed in the history of the Bulge, about $78\%$ are diverging systems. The majority ($\sim \! 80\%$) of present-day semi-detached systems (solid lines) have descended from converging LMXBs. All modeled present-day detached systems (dashed lines) have descended from diverging LMXBs, of which the donor envelope has been lost and the core has been left as a helium or carbon-oxygen white dwarf. The orbital periods of the white dwarf--neutron star systems are $\sim \! 6$ hr or longer. The shortest-period detached systems may have turned into UCXBs, if they have had enough time to shrink their orbit, and if they remained stable during the onset of mass transfer \citep{yungelson2002,vanhaaften2012evo}.


The orbital periods of the LMXBs vary between about $20$ min and $10$ d, with a peak around $2.5$ hr. Depending on donor mass, the bifurcation period is in the range $0.8 - 1.5$ d. The long-period end of the LMXB distribution represents diverging systems. The donor stars in present-day LMXBs in the Bulge originally had masses lower than $\sim \! 1\ M_{\odot}$.

\begin{figure}
\resizebox{\hsize}{!}{\includegraphics{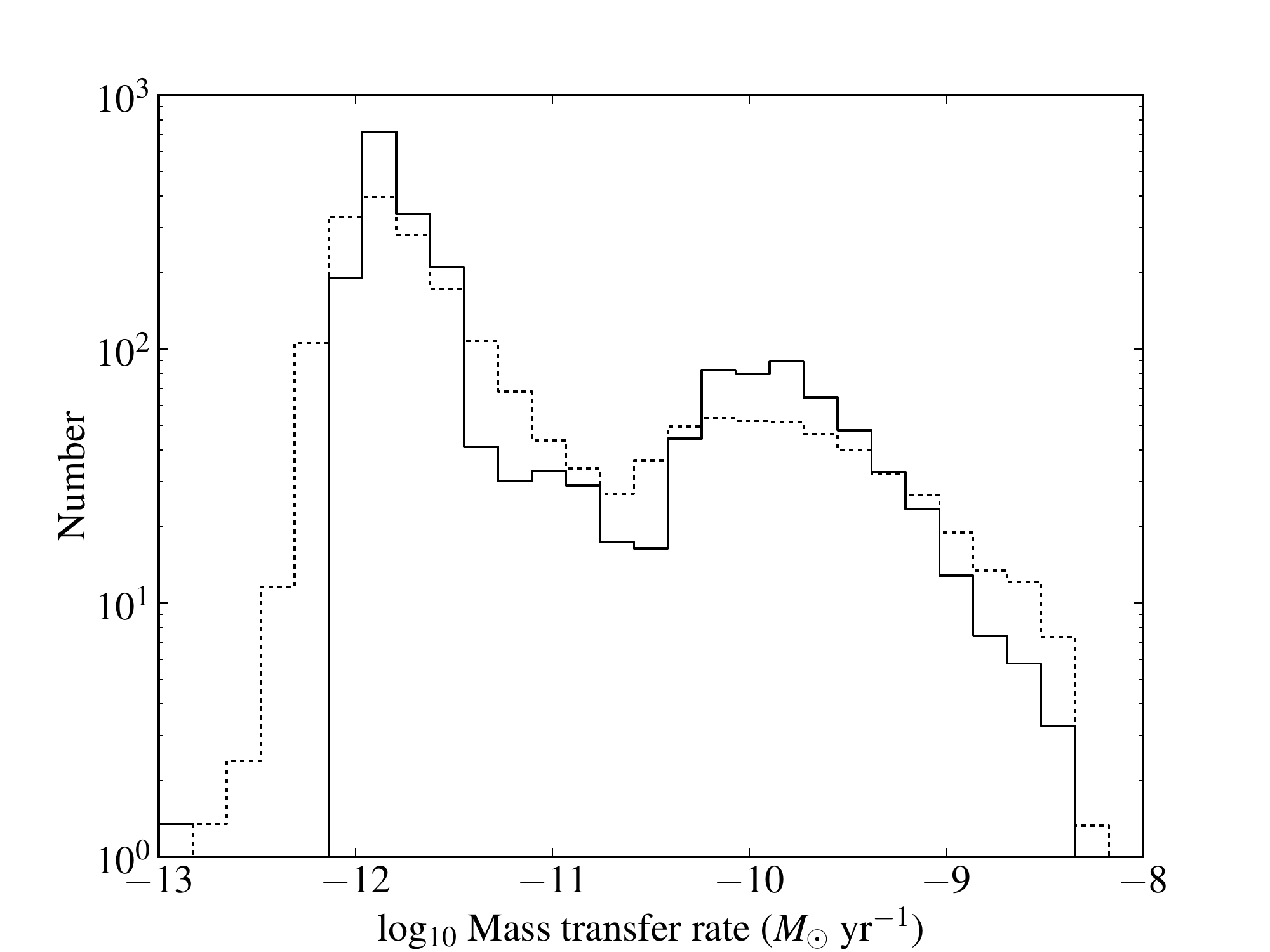}} 
\caption{Time-averaged mass transfer rate distributions of the present-day population of hydrogen-rich LMXBs with neutron star accretors in the Galactic Bulge. The star formation history width is $\sigma = 0.5$ Gyr (solid line) and $\sigma = 2.5$ Gyr (dotted line).}
\label{fig:lmxb:mdothist_log}
\end{figure}

Figure~\ref{fig:lmxb:mdothist_log} shows the time-averaged mass transfer rates of the neutron star LMXB population. About $330$ systems have time-averaged mass transfer rates exceeding $10^{-10}\ M_{\odot}\ \mathrm{yr}^{-1}$, approximately $1\%$ of the Eddington limit. The post-period minimum systems have lower average mass transfer rates than the population of diverging, long-period systems.

\subsection{The observable population of LMXBs}
\label{sect:lmxb:observable}

\begin{figure}
\resizebox{\hsize}{!}{\includegraphics{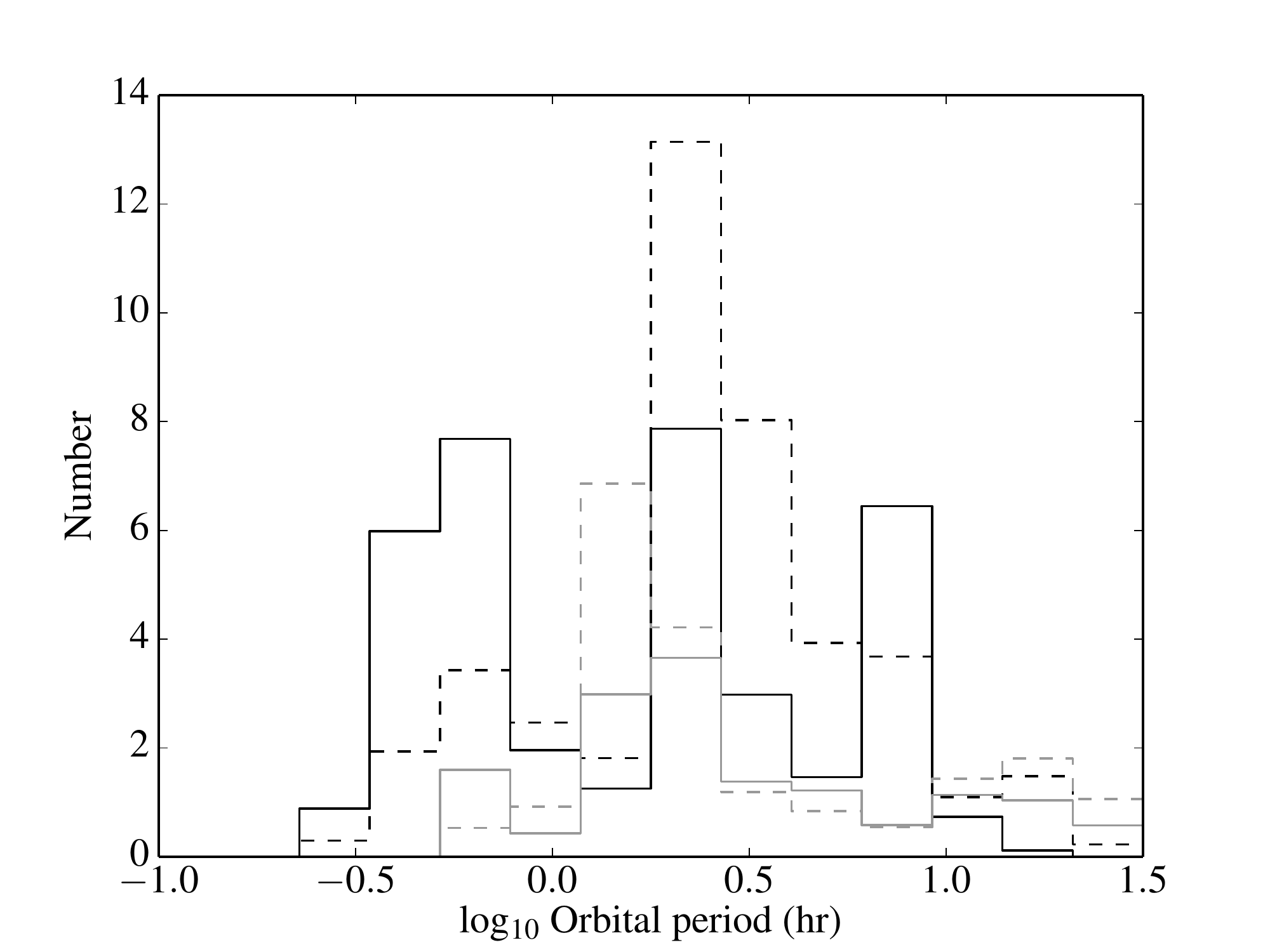}} 
\caption{Orbital period distributions of the present-day population of persistent hydrogen-rich LMXBs with neutron star accretors in the Galactic Bulge (black lines), as well as a snapshot in time of the transient population in outburst (gray lines). The star formation history width is $\sigma = 0.5$ Gyr (solid lines) and $\sigma = 2.5$ Gyr (dashed lines). Unlike the previous histograms, here the vertical scale is linear.}
\label{fig:lmxb:persist_per}
\end{figure}

\begin{figure}
\resizebox{\hsize}{!}{\includegraphics{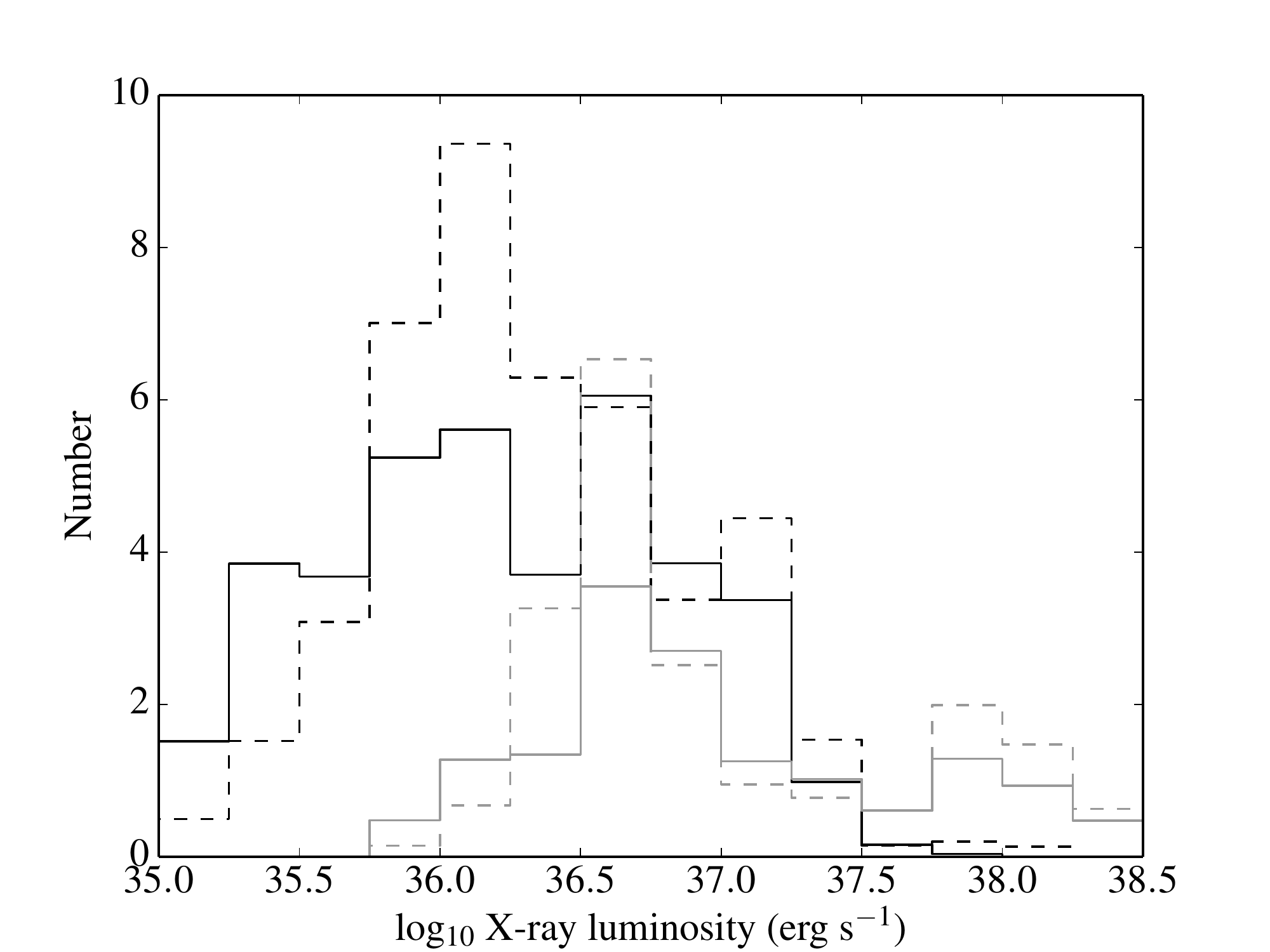}} 
\caption{X-ray luminosity distributions of the present-day population of persistent hydrogen-rich LMXBs with neutron star accretors in the Galactic Bulge (black lines), where mass transfer rates have been converted to X-ray luminosities using Eq.~(\ref{eq:lmxb:lum}), as well as a snapshot in time of the transient population in outburst (gray lines). The star formation history width is $\sigma = 0.5$ Gyr (solid lines) and $\sigma = 2.5$ Gyr (dashed lines).}
\label{fig:lmxb:persist_lum}
\end{figure}

As discussed in Sect.~\ref{sect:lmxb:method}, only the brightest and shortest-period sources in the modeled population are persistent, while the remainder of the population experience outbursts, with a duty cycle that decreases towards longer orbital period and lower average mass transfer rate. At the present time, the number of LMXBs in the persistent regime is $37$ for $\sigma = 0.5$ Gyr and $42$ for $\sigma = 2.5$ Gyr, whereas the number of transient LMXBs in outburst is $15$ for $\sigma = 0.5$ Gyr and $20$ for $\sigma = 2.5$ Gyr.
Figures~\ref{fig:lmxb:persist_per} and \ref{fig:lmxb:persist_lum} show that almost all persistent systems have orbital periods in the range $0.3 - 10$ hr and X-ray luminosities of $10^{35-37.5}\ \mbox{erg s}^{-1}$.
The transient sources in outburst are brighter on average than the persistent sources. The orbital periods of transients in outburst lie in the range $0.5 - 30$ hr, with a peak near $2$ hr, and their luminosities lie between $10^{36}\ \mbox{erg s}^{-1}$ and the Eddington luminosity.
The present-day observable population, both the persistent and transient sources, are predominantly of the converging type, especially for a narrow star formation history ($\sigma = 0.5$ Gyr).
The transient sources have mostly evolved beyond the period minimum. These correspond to the darkest squares in Figs.~\ref{fig:lmxb:density05} and \ref{fig:lmxb:density25}. The persistent sources are typically near, or on their way to, the period minimum.


It is interesting to note that the period distribution of the bright, observable population (Fig.~\ref{fig:lmxb:persist_per}) does not show a global maximum near the period minimum. Instead, the distribution is relatively uniform between $20$ min and $10$ hr. This is similar to the situation for Cataclysmic Variables, where a peak near the period minimum appears only after including very faint sources \citep{gansicke2009}.
In the LMXB case, our models predict an accumulation of semi-detached systems beyond the period minimum near $2$ hr (solid lines in Figs.~\ref{fig:lmxb:perhist_log05} and \ref{fig:lmxb:perhist_log25}), where angular momentum loss via gravitational wave radiation becomes very weak, and evolutionary timescales long. However, the vast majority of these systems is transient as a result of their low mass transfer rates. Because of their low duty cycles, only a small fraction is in outburst at a given time. On the other hand, systems with periods near $3 - 10$ hr are either persistent, or transient with relatively high duty cycles. Therefore there is no peak in the observable population at short orbital periods.

\section{Discussion}
\label{sect:lmxb:discussion}

We predict the existence of $\sim \! 40$ persistent LMXBs and, at any given time, $\sim \! 20$ transient LMXBs in outburst in the Galactic Bulge. These numbers are not very sensitive to the shape of the star formation history function.

However, the numbers of predicted persistent LMXBs (Sect.~\ref{sect:lmxb:observable}) are uncertain for several reasons. There are uncertainties in the population synthesis and LMXB tracks, but also in the disk irradiation \citep{dubus1999}. The latter is important because the critical lines in Figs.~\ref{fig:lmxb:density05} and \ref{fig:lmxb:density25} coincide with relatively narrow bands of LMXBs. For instance, increasing the critical mass transfer rate by a factor of $2$ would reduce the number of persistent sources by a factor of $2$ to $4$. A different critical mass transfer rate would also affect the duty cycle of transient sources.

If we use a lower value for the combined common-envelope parameter $\alpha_\mathrm{CE} \lambda = 0.2$ in \textsf{SeBa}, rather than the value of $2$ (Sect.~\ref{sect:lmxb:method}), as well as a Maxwellian velocity distribution of the neutron star kicks with a dispersion of $450\ \mbox{km s}^{-1}$, the total number of LMXBs formed is smaller by a factor of $\sim \! 2.7$. The relative numbers of persistent and transient systems do not change significantly.

Because the masses of neutron stars at birth are very uncertain, we adopted an initial neutron star mass of $1.4\ M_{\odot}$ in our simulations. However, \citet{kuranov2014} recently found that a uniform distribution of initial neutron star masses between $1.4 - 1.9\ M_{\odot}$ gives the best match with observations of persistent LMXBs in the Galactic Bulge.  
More massive neutron stars would affect our results in several ways. A more massive supernova remnant implies a higher probability for the binary system to survive the supernova event during which the neutron star is formed.
For a given Roche-lobe filling donor star, a higher accretor mass will cause a higher time-averaged mass transfer rate if mass transfer is driven by angular momentum loss via gravitational wave radiation ($\dot{M} \propto M_\mathrm{a}^{2/3}$, \citealp{vanhaaften2012evo}). Additionally, a more massive accretor and a higher mass transfer rate both directly lead to a higher average luminosity (see Eq.~\ref{eq:lmxb:lum}, where the neutron star mass-radius relation should also be taken into account).
The critical mass transfer rate (Eq.~\ref{eq:lmxb:inst}) also depends on accretor mass, although rather weakly ($\propto M_\mathrm{a}^{0.3}$).
Thus, the effect of a more massive accretor is as follows. Systems that are already persistent will become brighter. Transient systems close to the critical mass transfer rate may become persistent, however, persistent systems in which angular momentum loss is caused by magnetic braking may become transient. Transient systems well below the critical line will remain transient, but have a higher duty cycle.
A larger spread in neutron star masses will lead to a larger spread in observed X-ray luminosities as well.

Several studies \citep{pfahl2002,podsiadlowski2004} have suggested the existence of a dichotomy in neutron star kicks, where the kick velocity (and neutron star mass) depends on the type of supernova the neutron star is formed in, which in turn depends on the degree to which the primary was influenced by the companion during its evolution.
In our standard simulation, about $43\%$ of the neutron stars that end up in LMXBs originated from $8 - 11\ M_{\odot}$ primaries, which is a necessary condition for the formation of an electron capture supernova \citep{podsiadlowski2004}. Neutron stars forming in these supernovae are expected to have very low kick velocities ($\lesssim 50\ \mbox{km s}^{-1}$). Out of all neutron stars in LMXBs in our simulation, about $26\%$ went through a helium star/helium giant phase where the initial helium star mass was in the range of $2.6 - 2.95\ M_{\odot}$, which according to calculations by \citet{tauris2015} can lead to an electron capture supernova, depending on the initial orbital period.
The \citet{paczynski1990} kick velocity distribution, which we use in our standard model, includes a sufficient fraction of low kick velocities to account for the neutron stars formed during electron capture supernovae. Another reason we use the Paczy{\'n}ski kick distribution is to be able to directly compare the results to our earlier work on UCXBs \citep{vanhaaften2013bulgeucxb}, where we used the same distribution.

Several population synthesis studies of LMXBs have been performed in the past.
\citet{fragos2008} predicted $\sim \! 800$ LMXBs with neutron star accretors with $L_\mathrm{X} > 10^{36}\ \mbox{erg s}^{-1}$ in an elliptical galaxy (with an age similar to the Bulge), most of which are persistent. The stellar mass of the modeled elliptical galaxy is nine times larger than the stellar mass we use for the Bulge, therefore their estimate is equivalent to $\sim \! 90$ LMXBs in the Bulge. We find a total of $23 - 31$ persistent LMXBs with luminosities above this value in the Bulge, and at any given time another $15 - 20$ in outburst (for $\sigma = 0.5$ and $2.5$ Gyr, respectively), see Fig.~\ref{fig:lmxb:persist_lum}. Hence, our number of sources that is brighter than $10^{36}\ \mbox{erg s}^{-1}$ is a factor of $\sim \! 2 - 3$ smaller, per unit stellar mass. \citet{fragos2008} found most LMXBs near a period of $\sim \! 1$ hr, with a lower peak near $30$ hr. This resembles the distribution in our Figs.~\ref{fig:lmxb:perhist_log05} and \ref{fig:lmxb:perhist_log25} (solid lines), which show peaks near $2$ and $15$ hr.
\citet{kalogera1998two} and \citet{kalogera1998three} found a formation rate of LMXBs with neutron star accretors of $10^{-6} - 10^{-5}\ \mathrm{yr}^{-1}$ in the Galaxy, depending on common-envelope parameters and supernova kick velocities. This implies a total number of LMXBs formed over the history of the Disk of roughly $10^{4-5}$, which compares reasonably well to our result of $\sim \! 10^{4}$, taking into account that the Bulge has a stellar mass that is $4 - 6$ times lower than the stellar mass of the Disk \citep{klypin2002}.
\citet{kiel2006}, in their favored model, found about $1700$ LMXBs with neutron star accretors in the Galaxy.
\citet{kiel2013} modeled the population of LMXBs and black widow systems. Their various models produce $5 - 200$ black widow systems, and about one hundred times as many LMXBs in the Galaxy. These numbers are consistent with our results as well, again after correcting for the mass difference between Bulge and Galaxy.
\citet{kuranov2014} modeled an old stellar population of $500\,000$ systems. This corresponds to a total stellar mass of approximately $5 \times 10^{7} - 10^{8}\ M_{\odot}$, based on the stellar initial mass function. Therefore the numbers these authors predict, about $0.1 - 1$ LMXBs in their population across their models, should be multiplied by $100 - 200$ in order to compare them to our numbers for the Galactic Bulge. The number of LMXBs they found adjusted this way is a factor of $0.3 - 2$ times our number, depending on the models used. Their LMXBs are generally somewhat brighter than the systems we modeled, in particular they find more systems with luminosities over $10^{37}\ \mbox{erg s}^{-1}$.
We note that agreement between different models within a factor of a few is still acceptable, given for instance the uncertainty of a factor of $\sim \! 3$ based on the common-envelope parameters and kick velocity distribution.

The model results must be compared to the observed population of LMXBs, in particular bright systems. In general this is difficult because of the variability of LMXBs.
\citet{voss2010} found $18$ LMXBs in hard X-rays ($15 - 55$ keV) in data from the \textit{Swift} Burst Alert Telescope, within $10^{\circ}$ ($1.5$ kpc) of the Galactic Center, that are not known to be located outside the Bulge (e.g., in a globular cluster). Owing to confusion with background sources near the Galactic Center, there may be more sources. On the other hand, as a result of poor distance constraints, some of the observed sources may be located in front of, or behind, the Bulge. The faintest of the $18$ LMXBs has a luminosity of $\sim \! 2.4 \times 10^{35}\ \mbox{erg s}^{-1}$ in the hard X-ray band, which implies a total X-ray luminosity of $\sim 7 \times 10^{35}\ \mbox{erg s}^{-1}$ after a spectral correction based on \citet{revnivtsev2008}. The number of $18$ can be compared to our prediction in Fig.~\ref{fig:lmxb:persist_lum}, which shows about $45$ sources above this luminosity threshold.
The LMXB catalog by \citet{liu2007} contains approximately $40 - 50$ LMXBs (persistent and transient sources combined) that are located in the direction of the Bulge based on their celestial coordinates. Furthermore these sources are not known to be located in globular clusters, neither are they candidates of having black hole accretors. About one-third of these sources have a distance estimate that is consistent with the distance to, and size of, the Bulge. The others are located sufficiently close to the Galactic Center (e.g., $60\%$ within two degrees) that it is probable that the vast majority of them is located in the Bulge, given the high stellar density near the Galactic Center. The \citet{liu2007} catalog includes sources found by pointed observations, therefore the sample is inhomogeneous.
These numbers lie within a factor of two of our prediction of bright X-ray sources in the Bulge. A lower value for $\alpha_\mathrm{CE} \lambda$, as discussed above, would give an even better match. We consider the agreement of the number of predicted systems with the observed ones good.
We compare the orbital periods and luminosities of our modeled systems to persistent Galactic LMXBs in \citet{revnivtsev2011}. The X-ray luminosities of those systems in the $2 - 10$ keV range are $\sim 10^{36.5-38}\ \mbox{erg s}^{-1}$ at orbital periods of $\sim 2 - 5$ hr. This corresponds well with the properties of our persistent LMXBs, although we also find fainter systems with luminosities of $10^{35.5-36.5}\ \mbox{erg s}^{-1}$.
The Galactic Bulge Survey, which mostly aims to detect quiescent LMXBs \citep{jonker2011,jonker2014}, has observed $12$ square degrees in the Bulge. Extrapolating from our prediction of $\sim \! 60$ bright hydrogen-rich LMXBs in the Bulge as well as $\sim \! 50$ UCXBs \citep{vanhaaften2013bulgeucxb}, and assuming they are evenly distributed over the projected area of the Bulge, this implies up to six bright sources in the Galactic Bulge Survey field.

\subsection{Comparison with the population of ultracompact X-ray binaries}

Ultracompact X-ray binaries (UCXBs) are hydrogen-deficient LMXBs, with compact donors and observed orbital periods shorter than $1$ hr \citep[e.g.,][]{nelemans2010b,vanhaaften2012evo,heinke2013}.
As mentioned in Sect.~\ref{sect:lmxb:intro}, UCXBs can also form via an evolved main-sequence donor, but more than $99\%$ of all UCXBs is expected to form after a second common envelope phase, when eventually the core of the secondary fills its Roche lobe \citep{vanhaaften2013bulgeucxb}.

\begin{figure}
\resizebox{\hsize}{!}{\includegraphics{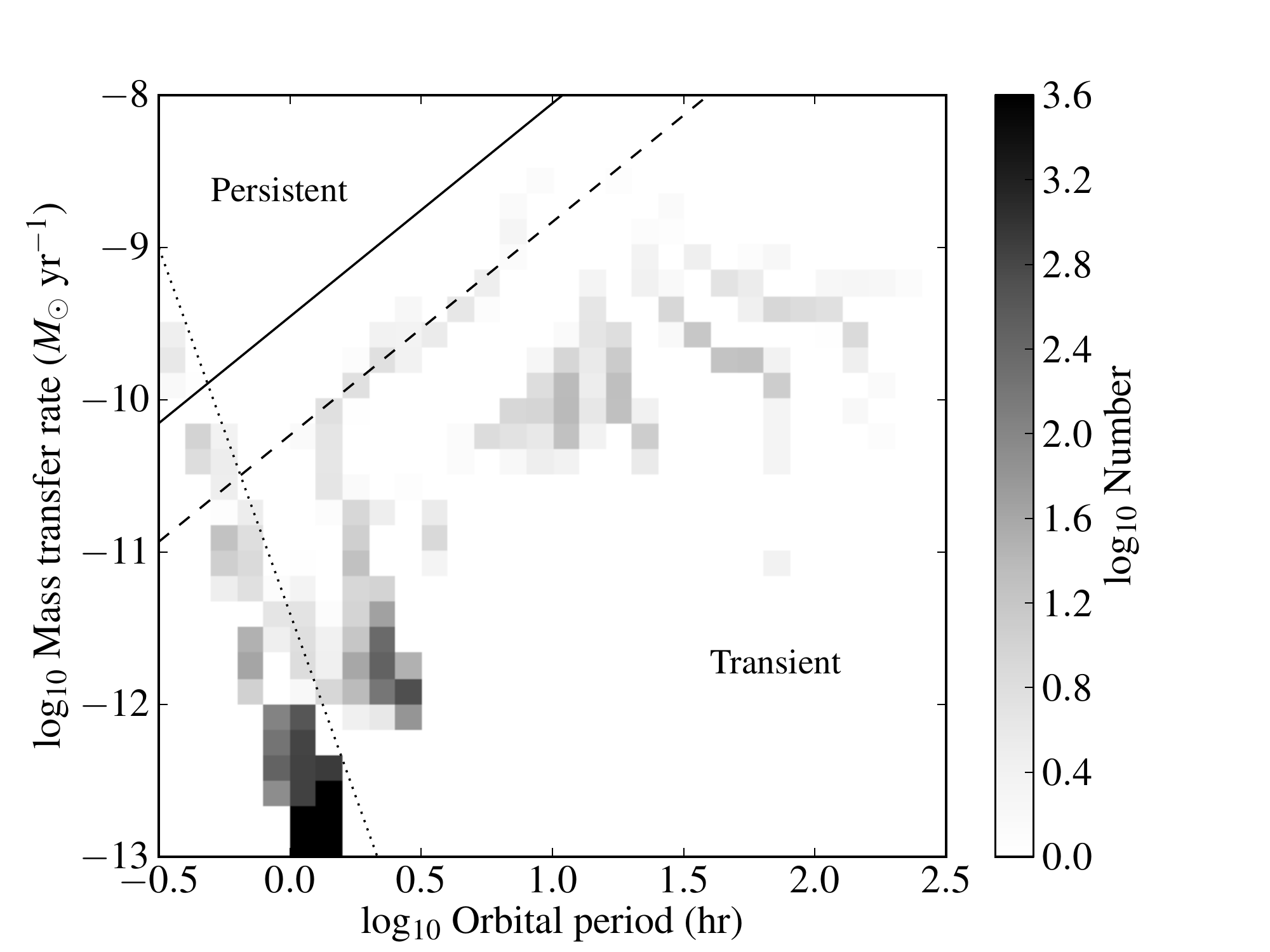}} 
\caption{Model distribution of present-day mass transfer rate versus orbital period for all UCXBs (the vast majority of systems to the left of the dotted line) and hydrogen-rich LMXBs (virtually all systems to the right of the dotted line) in the Galactic Bulge, assuming the UCXB evolution is driven only by gravitational wave radiation, and that all systems remain semi-detached. White squares correspond to one or fewer systems. The highest density in the figure is $10^{5}$ (outside the scale). The solid line shows the critical mass transfer rate for thermal-viscous disk instability, given by Eq.~(\ref{eq:lmxb:inst}) with $M_\mathrm{a} = 1.4\ M_{\odot}$ and $f = 6$ (helium composition), and likewise, the dashed line for $f = 1$ (solar composition). The star formation history width $\sigma = 0.5$ Gyr. The hydrogen-rich LMXB population in this figure is the same as in Fig.~\ref{fig:lmxb:density05}, but a different density scale is used. The UCXB data in this figure are taken from \citet{vanhaaften2013bulgeucxb}.}
\label{fig:lmxb:ucxbdensity05}
\end{figure}

Figure~\ref{fig:lmxb:ucxbdensity05} shows the present-day UCXB population in the Bulge (to the left of the dotted line) assuming their evolution is not affected by an irradiation-induced wind from the donor \citep{vanhaaften2013bulgeucxb}, as well as the population of hydrogen-rich LMXBs from Fig.~\ref{fig:lmxb:density05} (to the right of the dotted line) for comparison. The oldest UCXBs have the lowest mass transfer rates and the longest orbital periods. Without a donor wind, the evolution of an UCXB is driven by angular momentum loss only via gravitational wave radiation, and the system remains semi-detached. In the case of a wind from the donor, the mass transfer rate is higher, and longer orbital periods can be reached within the age of the Universe \citep{vanhaaften2012j1719}. Donor evaporation is expected to have an important impact on the orbital period distribution, which will shift towards longer periods and reduce the number of low mass transfer rate systems.

\citet{vanhaaften2013bulgeucxb} predicted a population of $(0.2 - 1.9) \times 10^{5}$ UCXBs in the Bulge, depending on model assumptions, of which a large fraction may have turned into (binary) millisecond pulsars as a result of recycling and accretion turnoff. When we compare the population of UCXBs in Fig.~\ref{fig:lmxb:ucxbdensity05} (showing $1.9 \times 10^{5}$ UCXBs) to the hydrogen-rich LMXBs ($2.1 \times 10^{3}$ systems), we see that UCXBs are about one hundred times more common than hydrogen-rich LMXBs, unless the vast majority of UCXBs have become detached and turned into millisecond radio pulsars (for the lower value of $\alpha_\mathrm{CE} \lambda = 0.2$ the ratio of $\sim \! 100$ decreases to $\sim \! 30$). In the case UCXBs typically become detached, hydrogen-rich LMXBs, such as the evaporating black widow systems, cannot be the main progenitor class of (isolated) millisecond radio pulsars. In fact, if most UCXBs become isolated millisecond radio pulsars, their absolute number corresponds reasonably well with the estimated number of isolated millisecond radio pulsars \citep{vanhaaften2013bulgeucxb}. Therefore, apart from the relative importance of UCXBs, the absolute numbers also suggest that the classical LMXBs cannot explain the size of the millisecond pulsar population.
The higher birth rate of UCXBs compared to hydrogen-rich LMXBs is probably a result of the higher secondary masses in zero-age main-sequence UCXB progenitor binaries, which make it easier to avoid disruption of the system during the supernova explosion of the primary, and to survive the common envelope stage.

If a significant fraction of UCXBs do not become millisecond radio pulsars, their number, $(0.2 - 1.9) \times 10^{5}$, is much higher than the number of hydrogen-rich LMXBs, $(0.8 - 2) \times 10^{3}$ (for the same variation in model parameters). However, the latter are expected to be slightly more common in the bright population ($\gtrsim 10^{35}\ \mbox{erg s}^{-1}$), with $20 - 60$ versus $5 - 50$ for UCXBs \citep{vanhaaften2013bulgeucxb}. Thus, a much larger fraction of hydrogen-rich LMXBs is bright, compared to UCXBs. This has several reasons. The donor composition in UCXBs can be helium, carbon and oxygen, or a mixture of these elements. LMXBs that start mass transfer from a hydrogen-rich donor have donors composed of hydrogen and/or helium. At low donor mass, the donors composed of lighter elements are larger than donors of heavier elements (of the same mass), due to Coulomb interactions \citep{zapolsky1969}. Therefore, the donors composed of heavier elements reach the same average donor density (hence also the same orbital period, by the period-density relation) at a lower donor mass. Because gravitational wave losses are smaller at lower donor mass, UCXBs have a lower time-averaged mass transfer rate at the same orbital period (and average density) than LMXBs with hydrogen in their donors (which can be seen in Fig.~\ref{fig:lmxb:ucxbdensity05}). Moreover, the critical mass transfer rates are higher for UCXBs (solid and dashed line in Fig.~\ref{fig:lmxb:ucxbdensity05}, respectively). Combined, this leads to significantly lower duty cycles for UCXBs -- this explains why the number of UCXBs in outburst is not larger than that of hydrogen-rich LMXBs in outburst, even though the UCXBs are much more common (unless they have become millisecond pulsars). Persistent sources have a higher critical mass transfer rate that results in a shorter life as a persistent source. The result is that a larger fraction of the population of hydrogen-rich LMXBs is persistent.

\begin{figure}
\resizebox{\hsize}{!}{\includegraphics{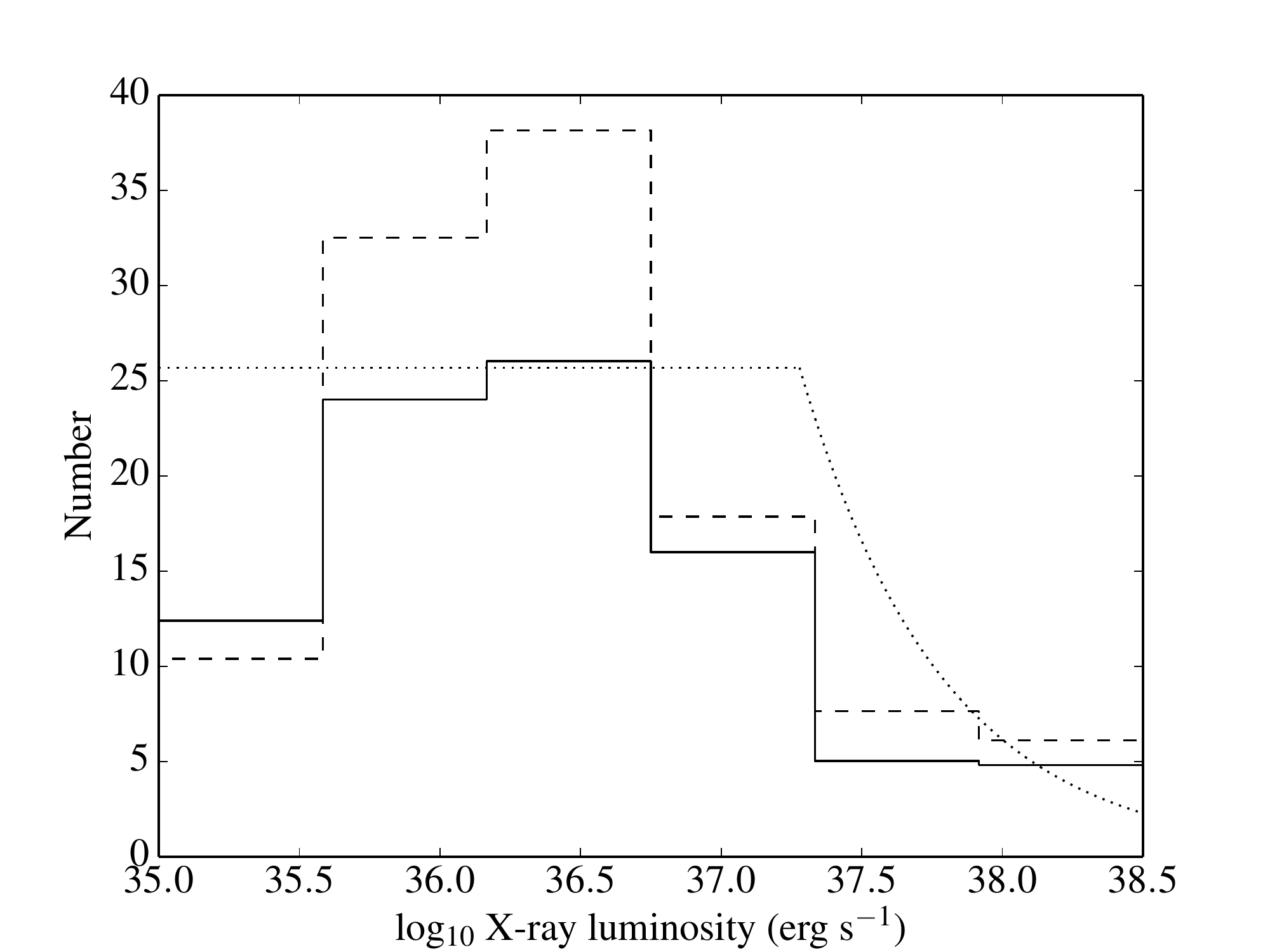}} 
\caption{X-ray luminosity distributions of the present-day population of LMXBs, including UCXBs, with neutron star accretors in the Galactic Bulge. The star formation history width is $\sigma = 0.5$ Gyr (solid line) and $\sigma = 2.5$ Gyr (dashed line). These distributions include hydrogen-rich as well as hydrogen-poor LMXBs, and both persistent sources and transient sources in outburst. The dotted line shows the X-ray luminosity function parametrized by \citet[][sample `All' in Table~3]{gilfanov2004} using the best-fitting normalization of $K_{1} = 440.4$, and further adjusted for the bin width of our histograms to allow for direct comparison.}
\label{fig:lmxb:lmxbucxb_lum}
\end{figure}

In Fig.~\ref{fig:lmxb:lmxbucxb_lum} we combine the luminosity distribution of hydrogen-rich LMXBs (Fig.~\ref{fig:lmxb:persist_lum}) with the luminosity distribution of UCXBs by \citet{vanhaaften2013bulgeucxb}, both the persistent sources and the transient sources in outburst, for our standard model ($\alpha_\mathrm{CE} \lambda = 2$). In this model, the contributions of persistent and transient systems are approximately equal, but in the case of a common envelope that is ten times less efficient, the number of UCXBs decreases by a factor of $\sim \! 8$, whereas the number of hydrogen-rich LMXBs decreases by a factor of $\sim \! 2.7$. Therefore the latter are more common among bright sources by a factor of a few with this change in assumptions.

The bright end of our luminosity distribution (Fig.~\ref{fig:lmxb:lmxbucxb_lum}) can be compared with extragalactic LMXB observations parametrized by \citet{gilfanov2004}, shown as the dotted curve in Fig.~\ref{fig:lmxb:lmxbucxb_lum}. The absolute numbers correspond quite well with our two models (solid and dashed bars). Counting only sources with luminosities between $10^{35} - 10^{38.5}\ \mbox{erg s}^{-1}$, our $\sigma = 0.5$ Gyr model has $90$ sources, and our $\sigma = 2.5$ Gyr model $110$. \citet{gilfanov2004} has $120$ sources in this luminosity range. Our alternative $\alpha_\mathrm{CE} \lambda = 0.2$ model produces about five times fewer sources than \citet{gilfanov2004}. The bright-end cut-off is located at a somewhat lower luminosity in our models, and is more gradual. Various uncertainties in the models could cause this. For instance, adopting a higher outburst luminosity for transient sources would make the distributions more similar. The decrease at low luminosities in our distribution is a result of not modeling quiescent stages, and in the case of transient sources, of assuming an abrupt transition from quiescence to outburst and back.

Figure~\ref{fig:lmxb:lmxbucxb_lum} also shows that our model including recent star formation ($\sigma = 2.5$ Gyr, histogram with dashed line) shows a stronger break near $10^{37}\ \mbox{erg s}^{-1}$ than the model without recent star formation (solid line). This suggests that particulars of the star formation history could be a factor in explaining why spiral galaxies like the \object{Milky Way} \citep{grimm2002} and \object{M~31} \citep{kong2003,trudolyubov2004} show a break near this luminosity, whereas this break is probably absent in the old elliptical galaxies \object{NGC~3379} and \object{NGC~4278} \citep{kim2006}.

\section{Conclusions}
\label{sect:lmxb:conclusion}

We have simulated the evolution of binaries in the Galactic Bulge in order to predict the present-day population of low-mass X-ray binaries and their possible descendants, such as detached white dwarf--neutron star systems and millisecond radio pulsars.
We predict about $2.0 \times 10^{3}$ LMXBs with neutron star accretors in the Galactic Bulge. Based on the predicted number of persistent sources and transient sources in outburst, this number corresponds well with observations. Because our prediction is supported by observations combined with the timescale progression along the evolutionary tracks for LMXBs (which are more accurate than population synthesis), it is not very sensitive to uncertainties in population synthesis models.
About three-quarters of the LMXBs are predicted to be systems below the bifurcation period that have passed through the period minimum, and have very low mass transfer rates and orbital periods shorter than $3$ hr, some of which may have become millisecond radio pulsars.
Most LMXBs that have formed over the history of the Bulge are no longer transferring mass at the present time, and we predict about $5.5 \times 10^{3}$ binary millisecond radio pulsars with white dwarf companions in the Bulge with a LMXB origin (i.e., excluding the contribution of intermediate-mass X-ray binaries).

Ultracompact X-ray binaries have a significantly higher formation rate than hydrogen-rich LMXBs and are the most common subclass of low-mass X-ray binaries, unless the majority have become detached and turned into millisecond radio pulsars, in which case they dominate the millisecond radio pulsar formation from all LMXBs. Even so, hydrogen-rich LMXBs are more easily observed because a larger fraction of them is persistent and the transient sources have longer duty cycles.

\begin{acknowledgements}
LMvH, GN, and RV are supported by the Netherlands Organisation for Scientific Research (NWO). GN and RV are supported by NWO Vidi grant $016.093.305$ to GN. This research has made use of NASA's Astrophysics Data System Bibliographic Services.
\end{acknowledgements}

\bibliographystyle{aa}
\bibliography{lennart_refs}

\begin{thebibliography}{65}
\expandafter\ifx\csname natexlab\endcsname\relax\def\natexlab#1{#1}\fi

\bibitem[{{Archibald} {et~al.}(2009){Archibald}, {Stairs}, {Ransom}, {Kaspi},
  {Kondratiev}, {Lorimer}, {McLaughlin}, {Boyles}, {Hessels}, {Lynch}, {van
  Leeuwen}, {Roberts}, {Jenet}, {Champion}, {Rosen}, {Barlow}, {Dunlap}, \&
  {Remillard}}]{archibald2009}
{Archibald}, A.~M., {Stairs}, I.~H., {Ransom}, S.~M., {et~al.} 2009, Science,
  324, 1411

\bibitem[{{Belczynski} {et~al.}(2008){Belczynski}, {Kalogera}, {Rasio}, {Taam},
  {Zezas}, {Bulik}, {Maccarone}, \& {Ivanova}}]{belczynski2008}
{Belczynski}, K., {Kalogera}, V., {Rasio}, F.~A., {et~al.} 2008, \apjs, 174,
  223

\bibitem[{{Clarkson} \& {Rich}(2009)}]{clarkson2009}
{Clarkson}, W. \& {Rich}, R.~M. 2009, in astro2010: The Astronomy and
  Astrophysics Decadal Survey, Science White Papers, 47

\bibitem[{{Dubus} {et~al.}(1999){Dubus}, {Lasota}, {Hameury}, \&
  {Charles}}]{dubus1999}
{Dubus}, G., {Lasota}, J., {Hameury}, J., \& {Charles}, P. 1999, \mnras, 303,
  139

\bibitem[{{Duquennoy} \& {Mayor}(1991)}]{duquennoy1991}
{Duquennoy}, A. \& {Mayor}, M. 1991, \aap, 248, 485

\bibitem[{{Fragos} {et~al.}(2008){Fragos}, {Kalogera}, {Belczynski},
  {Fabbiano}, {Kim}, {Brassington}, {Angelini}, {Davies}, {Gallagher}, {King},
  {Pellegrini}, {Trinchieri}, {Zepf}, {Kundu}, \& {Zezas}}]{fragos2008}
{Fragos}, T., {Kalogera}, V., {Belczynski}, K., {et~al.} 2008, \apj, 683, 346

\bibitem[{{Fruchter} {et~al.}(1988){Fruchter}, {Stinebring}, \&
  {Taylor}}]{fruchter1988}
{Fruchter}, A.~S., {Stinebring}, D.~R., \& {Taylor}, J.~H. 1988, \nat, 333, 237

\bibitem[{{G{\"a}nsicke} {et~al.}(2009){G{\"a}nsicke}, {Dillon}, {Southworth},
  {Thorstensen}, {Rodr{\'{\i}}guez-Gil}, {Aungwerojwit}, {Marsh}, {Szkody},
  {Barros}, {Casares}, {de Martino}, {Groot}, {Hakala}, {Kolb}, {Littlefair},
  {Mart{\'{\i}}nez-Pais}, {Nelemans}, \& {Schreiber}}]{gansicke2009}
{G{\"a}nsicke}, B.~T., {Dillon}, M., {Southworth}, J., {et~al.} 2009, \mnras,
  397, 2170

\bibitem[{{Gilfanov}(2004)}]{gilfanov2004}
{Gilfanov}, M. 2004, \mnras, 349, 146

\bibitem[{{Grimm} {et~al.}(2002){Grimm}, {Gilfanov}, \& {Sunyaev}}]{grimm2002}
{Grimm}, H.-J., {Gilfanov}, M., \& {Sunyaev}, R. 2002, \aap, 391, 923

\bibitem[{{Heinke} {et~al.}(2013){Heinke}, {Ivanova}, {Engel}, {Pavlovskii},
  {Sivakoff}, {Cartwright}, \& {Gladstone}}]{heinke2013}
{Heinke}, C.~O., {Ivanova}, N., {Engel}, M.~C., {et~al.} 2013, \apj, 768, 184

\bibitem[{{in't Zand} {et~al.}(2007){in't Zand}, {Jonker}, \&
  {Markwardt}}]{zand2007}
{in't Zand}, J.~J.~M., {Jonker}, P.~G., \& {Markwardt}, C.~B. 2007, \aap, 465,
  953

\bibitem[{{Jonker} {et~al.}(2011){Jonker}, {Bassa}, {Nelemans}, {Steeghs},
  {Torres}, {Maccarone}, {Hynes}, {Greiss}, {Clem}, {Dieball}, {Mikles},
  {Britt}, {Gossen}, {Collazzi}, {Wijnands}, {in't Zand}, {M{\'e}ndez}, {Rea},
  {Kuulkers}, {Ratti}, {van Haaften}, {Heinke}, {{\"O}zel}, {Groot}, \&
  {Verbunt}}]{jonker2011}
{Jonker}, P.~G., {Bassa}, C.~G., {Nelemans}, G., {et~al.} 2011, \apjs, 194, 18

\bibitem[{{Jonker} {et~al.}(2014){Jonker}, {Torres}, {Hynes}, {Maccarone},
  {Steeghs}, {Greiss}, {Britt}, {Wu}, {Johnson}, {Nelemans}, \&
  {Heinke}}]{jonker2014}
{Jonker}, P.~G., {Torres}, M.~A.~P., {Hynes}, R.~I., {et~al.} 2014, \apjs, 210,
  18

\bibitem[{{Joss} \& {Rappaport}(1979)}]{joss1979}
{Joss}, P.~C. \& {Rappaport}, S. 1979, \aap, 71, 217

\bibitem[{{Justham} {et~al.}(2006){Justham}, {Rappaport}, \&
  {Podsiadlowski}}]{justham2006}
{Justham}, S., {Rappaport}, S., \& {Podsiadlowski}, P. 2006, \mnras, 366, 1415

\bibitem[{{Kalogera}(1998)}]{kalogera1998three}
{Kalogera}, V. 1998, \apj, 493, 368

\bibitem[{{Kalogera}(1999)}]{kalogera1999}
{Kalogera}, V. 1999, \apj, 521, 723

\bibitem[{{Kalogera} \& {Webbink}(1998)}]{kalogera1998two}
{Kalogera}, V. \& {Webbink}, R.~F. 1998, \apj, 493, 351

\bibitem[{{Kiel} \& {Hurley}(2006)}]{kiel2006}
{Kiel}, P.~D. \& {Hurley}, J.~R. 2006, \mnras, 369, 1152

\bibitem[{{Kiel} \& {Taam}(2013)}]{kiel2013}
{Kiel}, P.~D. \& {Taam}, R.~E. 2013, \apss, 348, 441

\bibitem[{{Kim} {et~al.}(2006){Kim}, {Fabbiano}, {Kalogera}, {King},
  {Pellegrini}, {Trinchieri}, {Zepf}, {Zezas}, {Angelini}, {Davies}, \&
  {Gallagher}}]{kim2006}
{Kim}, D.-W., {Fabbiano}, G., {Kalogera}, V., {et~al.} 2006, \apj, 652, 1090

\bibitem[{{Klypin} {et~al.}(2002){Klypin}, {Zhao}, \&
  {Somerville}}]{klypin2002}
{Klypin}, A., {Zhao}, H., \& {Somerville}, R.~S. 2002, \apj, 573, 597

\bibitem[{{Kong} {et~al.}(2003){Kong}, {DiStefano}, {Garcia}, \&
  {Greiner}}]{kong2003}
{Kong}, A.~K.~H., {DiStefano}, R., {Garcia}, M.~R., \& {Greiner}, J. 2003,
  \apj, 585, 298

\bibitem[{{Kroupa}(2001)}]{kroupa2001}
{Kroupa}, P. 2001, \mnras, 322, 231

\bibitem[{{Kuranov} {et~al.}(2014){Kuranov}, {Postnov}, \&
  {Revnivtsev}}]{kuranov2014}
{Kuranov}, A.~G., {Postnov}, K.~A., \& {Revnivtsev}, M.~G. 2014, Astronomy
  Letters, 40, 29

\bibitem[{{Lasota}(2001)}]{lasota2001}
{Lasota}, J. 2001, \nar, 45, 449

\bibitem[{{Liu} {et~al.}(2007){Liu}, {van Paradijs}, \& {van den
  Heuvel}}]{liu2007}
{Liu}, Q.~Z., {van Paradijs}, J., \& {van den Heuvel}, E.~P.~J. 2007, \aap,
  469, 807

\bibitem[{{Nelemans} \& {Jonker}(2010)}]{nelemans2010b}
{Nelemans}, G. \& {Jonker}, P.~G. 2010, \nar, 54, 87

\bibitem[{{Nelemans} {et~al.}(2001){Nelemans}, {Yungelson}, {Portegies Zwart},
  \& {Verbunt}}]{nelemans2001a}
{Nelemans}, G., {Yungelson}, L.~R., {Portegies Zwart}, S.~F., \& {Verbunt}, F.
  2001, \aap, 365, 491

\bibitem[{{Nelson} \& {Rappaport}(2003)}]{nelson2003}
{Nelson}, L.~A. \& {Rappaport}, S. 2003, \apj, 598, 431

\bibitem[{{Paczy{\'n}ski}(1981)}]{paczynski1981cv}
{Paczy{\'n}ski}, B. 1981, \actaa, 31, 1

\bibitem[{{Paczy{\'n}ski}(1990)}]{paczynski1990}
{Paczy{\'n}ski}, B. 1990, \apj, 348, 485

\bibitem[{{Paczy{\'n}ski} \& {Sienkiewicz}(1981)}]{paczynski1981sien}
{Paczy{\'n}ski}, B. \& {Sienkiewicz}, R. 1981, \apjl, 248, L27

\bibitem[{{Pfahl} {et~al.}(2002){Pfahl}, {Rappaport}, {Podsiadlowski}, \&
  {Spruit}}]{pfahl2002}
{Pfahl}, E., {Rappaport}, S., {Podsiadlowski}, P., \& {Spruit}, H. 2002, \apj,
  574, 364

\bibitem[{{Podsiadlowski} {et~al.}(2004){Podsiadlowski}, {Langer},
  {Poelarends}, {Rappaport}, {Heger}, \& {Pfahl}}]{podsiadlowski2004}
{Podsiadlowski}, P., {Langer}, N., {Poelarends}, A.~J.~T., {et~al.} 2004, \apj,
  612, 1044

\bibitem[{{Podsiadlowski} {et~al.}(2003){Podsiadlowski}, {Rappaport}, \&
  {Han}}]{podsiadlowski2003}
{Podsiadlowski}, P., {Rappaport}, S., \& {Han}, Z. 2003, \mnras, 341, 385

\bibitem[{{Podsiadlowski} {et~al.}(2002){Podsiadlowski}, {Rappaport}, \&
  {Pfahl}}]{podsiadlowski2002}
{Podsiadlowski}, P., {Rappaport}, S., \& {Pfahl}, E.~D. 2002, \apj, 565, 1107

\bibitem[{{Popova} {et~al.}(1982){Popova}, {Tutukov}, \&
  {Yungelson}}]{popova1982}
{Popova}, E.~I., {Tutukov}, A.~V., \& {Yungelson}, L.~R. 1982, \apss, 88, 55

\bibitem[{{Portegies Zwart} \& {Verbunt}(1996)}]{portegieszwart1996}
{Portegies Zwart}, S.~F. \& {Verbunt}, F. 1996, \aap, 309, 179

\bibitem[{{Portegies Zwart} {et~al.}(1997){Portegies Zwart}, {Verbunt}, \&
  {Ergma}}]{portegieszwart1997}
{Portegies Zwart}, S.~F., {Verbunt}, F., \& {Ergma}, E. 1997, \aap, 321, 207

\bibitem[{{Portegies Zwart} \& {Yungelson}(1998)}]{portegieszwart1998}
{Portegies Zwart}, S.~F. \& {Yungelson}, L.~R. 1998, \aap, 332, 173

\bibitem[{{Pylyser} \& {Savonije}(1988)}]{pylyser1988}
{Pylyser}, E. \& {Savonije}, G.~J. 1988, \aap, 191, 57

\bibitem[{{Pylyser} \& {Savonije}(1989)}]{pylyser1989}
{Pylyser}, E.~H.~P. \& {Savonije}, G.~J. 1989, \aap, 208, 52

\bibitem[{{Rappaport} {et~al.}(1982){Rappaport}, {Joss}, \&
  {Webbink}}]{rappaport1982}
{Rappaport}, S., {Joss}, P.~C., \& {Webbink}, R.~F. 1982, \apj, 254, 616

\bibitem[{{Rappaport} {et~al.}(1983){Rappaport}, {Verbunt}, \&
  {Joss}}]{rappaport1983}
{Rappaport}, S., {Verbunt}, F., \& {Joss}, P.~C. 1983, \apj, 275, 713

\bibitem[{{Revnivtsev} {et~al.}(2008){Revnivtsev}, {Lutovinov}, {Churazov},
  {Sazonov}, {Gilfanov}, {Grebenev}, \& {Sunyaev}}]{revnivtsev2008}
{Revnivtsev}, M., {Lutovinov}, A., {Churazov}, E., {et~al.} 2008, \aap, 491,
  209

\bibitem[{{Revnivtsev} {et~al.}(2011){Revnivtsev}, {Postnov}, {Kuranov}, \&
  {Ritter}}]{revnivtsev2011}
{Revnivtsev}, M., {Postnov}, K., {Kuranov}, A., \& {Ritter}, H. 2011, \aap,
  526, A94

\bibitem[{{Ruderman} {et~al.}(1989{\natexlab{a}}){Ruderman}, {Shaham}, \&
  {Tavani}}]{ruderman1989}
{Ruderman}, M., {Shaham}, J., \& {Tavani}, M. 1989{\natexlab{a}}, \apj, 336,
  507

\bibitem[{{Ruderman} {et~al.}(1989{\natexlab{b}}){Ruderman}, {Shaham},
  {Tavani}, \& {Eichler}}]{ruderman1989late}
{Ruderman}, M., {Shaham}, J., {Tavani}, M., \& {Eichler}, D.
  1989{\natexlab{b}}, \apj, 343, 292

\bibitem[{{Tauris} {et~al.}(2015){Tauris}, {Langer}, \&
  {Podsiadlowski}}]{tauris2015}
{Tauris}, T.~M., {Langer}, N., \& {Podsiadlowski}, P. 2015, ArXiv e-prints
  [\eprint[arXiv]{1505.00270v1}]

\bibitem[{{Toonen} {et~al.}(2012){Toonen}, {Nelemans}, \& {Portegies
  Zwart}}]{toonen2012}
{Toonen}, S., {Nelemans}, G., \& {Portegies Zwart}, S.~F. 2012, \aap, 546, A70

\bibitem[{{Trudolyubov} \& {Priedhorsky}(2004)}]{trudolyubov2004}
{Trudolyubov}, S. \& {Priedhorsky}, W. 2004, \apj, 616, 821

\bibitem[{{Tutukov} {et~al.}(1985){Tutukov}, {Fedorova}, {Ergma}, \&
  {Yungelson}}]{tutukov1985}
{Tutukov}, A.~V., {Fedorova}, A.~V., {Ergma}, E.~V., \& {Yungelson}, L.~R.
  1985, Soviet Astronomy Letters, Vol.11, No.1/Jan-Feb, P.~52, 1985, 11, 52

\bibitem[{{van den Heuvel}(1975)}]{vandenheuvel1975}
{van den Heuvel}, E.~P.~J. 1975, \apjl, 198, L109

\bibitem[{{van der Sluys} {et~al.}(2005){van der Sluys}, {Verbunt}, \&
  {Pols}}]{sluys2005a}
{van der Sluys}, M.~V., {Verbunt}, F., \& {Pols}, O.~R. 2005, \aap, 431, 647

\bibitem[{{van Haaften} {et~al.}(2012{\natexlab{a}}){van Haaften}, {Nelemans},
  {Voss}, \& {Jonker}}]{vanhaaften2012j1719}
{van Haaften}, L.~M., {Nelemans}, G., {Voss}, R., \& {Jonker}, P.~G.
  2012{\natexlab{a}}, \aap, 541, A22

\bibitem[{{van Haaften} {et~al.}(2013){van Haaften}, {Nelemans}, {Voss},
  {Toonen}, {Portegies Zwart}, {Yungelson}, \& {van der
  Sluys}}]{vanhaaften2013bulgeucxb}
{van Haaften}, L.~M., {Nelemans}, G., {Voss}, R., {et~al.} 2013, \aap, 552, A69

\bibitem[{{van Haaften} {et~al.}(2012{\natexlab{b}}){van Haaften}, {Nelemans},
  {Voss}, {Wood}, \& {Kuijpers}}]{vanhaaften2012evo}
{van Haaften}, L.~M., {Nelemans}, G., {Voss}, R., {Wood}, M.~A., \& {Kuijpers},
  J. 2012{\natexlab{b}}, \aap, 537, A104

\bibitem[{{Voss} \& {Ajello}(2010)}]{voss2010}
{Voss}, R. \& {Ajello}, M. 2010, \apj, 721, 1843

\bibitem[{{Webbink} {et~al.}(1983){Webbink}, {Rappaport}, \&
  {Savonije}}]{webbink1983}
{Webbink}, R.~F., {Rappaport}, S., \& {Savonije}, G.~J. 1983, \apj, 270, 678

\bibitem[{{Wyse}(2009)}]{wyse2009}
{Wyse}, R.~F.~G. 2009, in IAU Symposium, Vol. 258, IAU Symposium, ed.
  {E.~E.~Mamajek, D.~R.~Soderblom, \& R.~F.~G.~Wyse}, 11--22

\bibitem[{{Yungelson} {et~al.}(2006){Yungelson}, {Lasota}, {Nelemans}, {Dubus},
  {van den Heuvel}, {Dewi}, \& {Portegies Zwart}}]{yungelson2006}
{Yungelson}, L.~R., {Lasota}, J., {Nelemans}, G., {et~al.} 2006, \aap, 454, 559

\bibitem[{{Yungelson} {et~al.}(2002){Yungelson}, {Nelemans}, \& {van den
  Heuvel}}]{yungelson2002}
{Yungelson}, L.~R., {Nelemans}, G., \& {van den Heuvel}, E.~P.~J. 2002, \aap,
  388, 546

\bibitem[{{Zapolsky} \& {Salpeter}(1969)}]{zapolsky1969}
{Zapolsky}, H.~S. \& {Salpeter}, E.~E. 1969, \apj, 158, 809

\end{thebibliography}

\end{document}